\documentclass[aip,jcp]{revtex4-1}

\usepackage{amssymb}
\usepackage{graphicx}% Include figure files
\usepackage{dcolumn}% Align table columns on decimal point
\usepackage{bm}% bold math
\usepackage{longtable}

\begin{document}

\newcommand{\cm}{cm$^{-1}$} \newcommand{\A}{\AA$^{-1}$} \newcommand{\Q}{\mathbf{Q}}
\newcommand{\PP}{\mathcal{P}} \newcommand{\TKHS}{K$_3$H(SO$_4)_2$} \newcommand{\TKDS}{K$_3$D(SO$_4)_2$} \newcommand{\TRbHS}{Rb$_3$H(SO$_4)_2$}
\newcommand{\TRbDS}{Rb$_3$D(SO$_4)_2$}

\draft

\title{Neutron scattering studies of K$_3$H(SO$_4)_2$ and K$_3$D(SO$_4)_2$: The particle-in-a-box model for the quantum phase transition }

\author{Fran\c{c}ois Fillaux}
\email{francois.fillaux@upmc.fr}
\affiliation{UMR 7075, CNRS,
Universit\'{e} Pierre et Marie Curie, Box 49,
4 place Jussieu, 75252 PARIS Cedex 05, France}
\author{Alain Cousson}
\affiliation{Laboratoire L\'{e}on Brillouin (CEA-CNRS), C.E. Saclay, 91191
Gif-sur-Yvette, cedex, France}

\date{\today}

\begin{abstract}
In the crystal of K$_3$H(SO$_4)_2$, or K$_3$D(SO$_4)_2$, dimers (SO$_4)\cdots$H$\cdots$(SO$_4)$, or (SO$_4)\cdots$D$\cdots$(SO$_4)$, are linked by strong centrosymmetric hydrogen or deuterium bonds whose O$\cdots$O length is $\approx 2.50$ \AA. We address two open questions. (i) Are H or D sites split or not? (ii) Is there any structural counterpart to the phase transition observed for K$_3$D(SO$_4)_2$ at $T_c \approx 85.5$ K, that does not exist for K$_3$H(SO$_4)_2$. Neutron diffraction by single-crystals at cryogenic or room temperature reveals no structural transition and no resolvable splitting of H or D sites. However, the width of the probability densities suggest unresolved splitting of the wavefunctions suggesting rigid entities H$_{L1/2}-$H$_{R1/2}$ or D$_{L1/2}-$D$_{R1/2}$ whose separation lengths are $l_\mathrm{H} \approx 0.16$ \AA\ or $l_\mathrm{D} \approx 0.25$ \AA. The vibrational eigenstates for the center of mass of H$_{L1/2}-$H$_{R1/2}$ revealed by inelastic neutron scattering are amenable to a square-well and we suppose the same potential holds for D$_{L1/2}-$D$_{R1/2}$. In order to explain dielectric and calorimetric measurements of mixed crystals K$_3$D$_{(1-\rho)}$H$_\rho$(SO$_4)_2$ ($0 \le \rho \le 1$), we replace the classical notion of order-disorder by the quantum notion of discernible (eg D$_{L1/2}-$D$_{R1/2}$) or indiscernible (eg H$_{L1/2}-$H$_{R1/2}$) components depending on the separation length of the split wavefunction. The discernible-indiscernible isostructural transition at finite temperatures is induced by a thermal pure quantum state or at 0 K by $\rho$.
\end{abstract}

\maketitle

\section{Introduction}

Symmetric hydrogen bonds O$\cdots$H$\cdots$O are of fundamental importance to many disciplines across physics, chemistry, or biology, in the gaseous, liquid or solid state. Their properties depend on essentially the O$\cdots$O bond length (say $R_\mathrm{{OO}}$), such that a shorter $R_\mathrm{{OO}}$ corresponds to a stronger bonding and a lower frequency for the OH stretching mode ($\nu$OH). This dependence on $R_\mathrm{{OO}}$ accounts for temperature, pressure, and isotope effects. As $R_\mathrm{OO}$ decreases, the potential operator for the stretching coordinate is expected to evolve from a double-well with a high barrier (compared to $kT$) to a single-well for the shortest bonds. The latter is widely thought of as mimicking symmetric intermediates for proton transfer along the $\nu$OH coordinate in (bio)chemical reactions. This is an important field of investigations for computational chemistry and our purpose is to determine effective potentials based upon experiments for benchmark tests of theoretical methods.

In crystals, strong symmetric hydrogen bonds are encountered for small entities, such as chelates or dimers, well separated from each other. Bond lengths, symmetry and probability densities at proton sites are best characterized with single-crystal neutron diffraction (SXND), compared to x-ray (SXXD). Inelastic neutron scattering (INS) reveals OH transitions with spectacular contrast of intensity, compared to infrared or Raman. Probability densities in the momentum space can be determined with neutron Compton scattering (NCS).

The best characterized short symmetric hydrogen bond in a crystal is that of the chelate of the potassiumhydrogenmaleate, KH(OOC-CH=CH-COO), for which $R_\mathrm{OO} = 2.427(1)$ \AA\ at 20 K. \cite{FLTCP} This is one of the shortest distance ever reported. INS spectra reveal numerous $\nu$OH transitions consistent with a funnel-shaped potential, such that the bare proton is localized at the center in the ground state, or largely delocalized over the oxygen orbitals in the excited states beyond $\approx 500$ \cm. Similar potentials have been proposed for centrosymmetric dimers of potassiumhydrogenbistrifluoroacetate, KH(CF$_3$COO)$_2$ and the cesium analogue, CsH(CF$_3$COO)$_2$, both with $R_\mathrm{OO} = 2.436(4)$ \AA\ at cryogenic temperatures. \cite{FCAT} In crystals, the existence of symmetric double-wells for longer bonds is controversial. For the crystal of potassiumdihydrogenphosphate (KDP, $R_\mathrm{OO} \approx 2.50$ \AA), Reiter et al. \cite{RMP} claimed that NCS suggests a symmetric double-well in the paraelectric phase above $T_c \approx 124$ K, or a single minimum in the ferroelectric phase below $T_c$. However, this is clearly excluded by the crystal symmetry and vibrational spectra. \cite{TKA} Quasi-symmetric double wells have been reported for centrosymmetric dimers of the hydrogencarbonate family ($M$HCO$_3$ with $M =$ K, Rb, Cs), \cite{Fil2,FT1} or benzoic acid, \cite{FLR,FRLL} for which $R_\mathrm{OO} \approx 2.60$ \AA, but these hydrogen bonds are not symmetric. Apart from crystals, the chelate of the malonaldehyde (3-hydroxy-2-propenal) molecule in the gas phase ($R_\mathrm{{OO}} \approx 2.57$ \AA) is the best characterized symmetric double-well. \cite{BTMYT,FN}

In the present work we report experimental and theoretical studies of the crystals of trispotassiumhydrogendisulfate, K$_3$H(SO$_4)_2$, and of the isostructural \TKDS. \cite{NUKKT,NKWT,SR} The unit cell of \TKHS\ is comprised of 4 equivalent centrosymmetric dimers, SO$_4\cdots$H$\cdots$SO$_4$ for which $R_\mathrm{OO} \approx 2.515(2)$ \AA\ at 293 K or $\approx 2.487(2)$ \AA\ at 100 K. These small dimers are well separate by K nuclei. The monoclinic group symmetry is $A2/a$ and there is no structural phase-transition detectable by SXXD from 4 K to 300 K. Noda et al. \cite{NKWT} suggested that split proton sites at room temperature turn into single proton sites at low temperature, as $R_\mathrm{OO}$ becomes shorter (geometric effect). INS reveals OH bending modes at 1250 or 1600 \cm\ and the $\nu$OH mode at $\approx 57$ \cm. This is the lowest frequency ever reported for any hydrogen bond in a crystal. \cite{FLTK} This could be an evidence to suggest an asymmetric double-well with a low barrier, but this is forbidden by the crystal symmetry. We report below SXND measurements showing that a double-well is unlikely at every temperature and we propose a new potential operator for INS.

Several members of the $M_3$H($X$O$_4)_2$ family ($M =$ K, Rb, Cs; $X =$ S, Se) demonstrate remarkable isotope effects which reveal intriguing differences between H and D bonds. In the specific case of \TKDS, calorimetric and dielectric measurements reveal anomalies attributed to a phase transition at $T_c \approx 85.5$ K, whereas there is no transition for \TKHS. \cite{MTNSK} So far, $T_c$ is conceived of as an order-disorder transition from an antiferroelectric phase ($A2$) below $T_c$ to a paraelectric phase ($A2/a$) above $T_c$, \cite{Gesi1,MTNSK,NTNMK} A widespread explanation is that deuterons are delocalized in a symmetric double-well above $T_c$ and statistically localized below $T_c$. \cite{NUKKT,NKWTG,NKWT,NTNMK,DLMST,*DLMSSP,*DLMSGP} Two models have been proposed to explain the isotope effect. (i) The pseudospin Ising model with markedly different tunnel splitting for H or D. \cite{MTNSK,NTNMK,DLMST,DLMSSP,DLMSGP} (ii) Markedly different geometric effects for H or D bonds. \cite{MM,NKWTG,NUKKT,NKWT} Both models are weakened by the lack of unquestionable evidence to suggest $A2$ symmetry consistent with asymmetric O$\cdots$D$\cdots$O bonds below $T_c$. We have, therefore, carried out SXND measurements in order to observe whether or not there is a structural counterpart (possibly $A2 \longleftrightarrow A2/a$) to the dielectric and calorimetric anomalies.

This article is organized as follows. In Sec. \ref{sec:2}, we present SXND data for \TKHS\ and \TKDS\ at cryogenic and room temperatures. It transpires that the space group is $A2/a$ at every temperature for both crystals and the best refined structures are obtained with H or D at the center of symmetry. Statistical distributions and antiferroelectricity below $T_c$ are, therefore, ruled out. In Sec. \ref{sec:3}, we show that SXND and INS are together consistent with unresolved split protons in a square-well and the same model is supposed to hold for deuterons. In Sec. \ref{sec:4} we propose a new interpretation within the framework of quantum mechanics of the dielectric and calorimetric anomalies for KD$_{(1-\rho)}$H$_\rho$(SO$_4)_2$, $0 \le \rho \le 1$. We emphasize that these crystals are macroscopically quantum, to the least, below $T_c$. In Sec. \ref{sec:5} we show that the same model holds for the isostructural \TRbHS.

\section{\label{sec:2}Neutron diffraction}

Single-crystals were obtained by slow cooling of H$_2$SO$_4$(D$_2$SO$_4$);$1.2 $ K$_2$SO$_4$  in H$_2$O(D$_2$O). Prismatic colorless specimens wrapped in aluminum were loaded in a closed-cycle-refrigerator whose temperature was controlled to $\pm 1$ K. Data were collected with the four-circle diffractometer 5C2 at the Orph\'{e}e reactor (Laboratoire L\'{e}on Brillouin). \cite{LLB} Every structural parameter computed with CRYSTALS \cite{CRYSTALS} was allowed to vary independently. \cite{SUPM}

\begin{figure}[ht]
\includegraphics[angle=0.,scale=0.6]{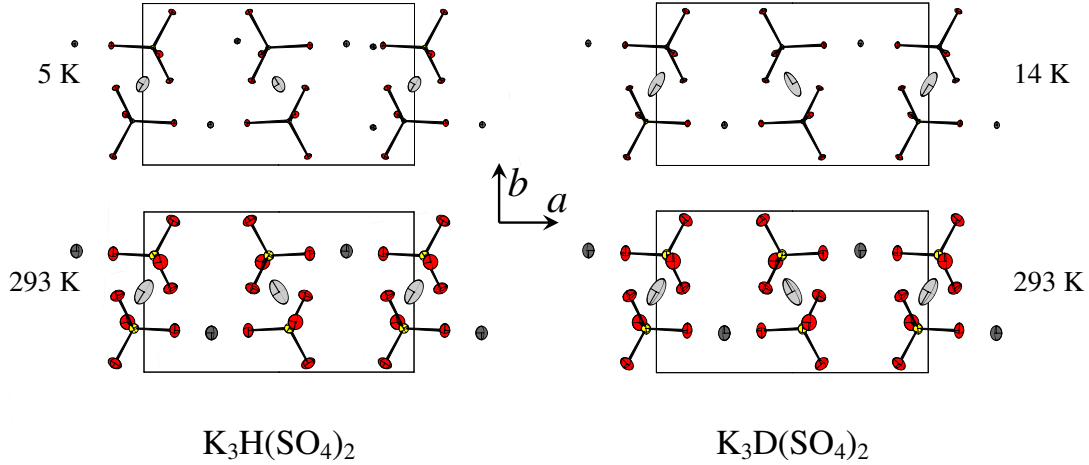}
\caption{\label{fig:abHD} Projections of dimer entities onto $(a,b)$ planes. Thermal ellipsoids represent $50 \%$ of the probability density for nuclei. Sulfur is yellow, oxygen is red, potassium is black, hydrogen or deuterium are grey. }
\end{figure}

Inspection of intensities for absent reflections confirms the monoclinic space group $A2/a$ for both K$_3$H(SO$_4)_2$ and K$_3$D(SO$_4)_2$ at every temperature (see Table \ref{tab:1}). There is no visible symmetry breaking to suggest a statistical distribution of H or D. Best refinements were obtained with H or D at special positions (Tables \ref{tab:2} and \ref{tab:3}) and every attempt to split these sites was less satisfactory. The best refined structures are comprised of Z = 4 dimer entities per unit cell. The slightly distorted SO$_4$ tetrahedra are consistent with spontaneous dielectric polarization. They are linked through centrosymmetric O$\cdots$H$\cdots$O or O$\cdots$D$\cdots$O bonds practically parallel to $(a,b)$ and at $\approx\pm 30^\circ$ with respect to $b$ (Fig. \ref{fig:abHD}). The supposedly antiferroelectric phase of \TKDS\ below $T_c$ is not confirmed and both the Ising model and geometric effects are rejected.

Unit cell parameters (Table \ref{tab:1}), positional parameters (Tables \ref{tab:2} and \ref{tab:3})), and thermal parameters for heavy nuclei (Tables \ref{tab:4} and \ref{tab:5}) are similar for both isotopomers but temperature effects are different. For \TKHS, the volume of the unit cell increases by $\approx 3\%$ between 5 K and 293 K, $R_{\mathrm{OHO}}$ increases from 2.463(1) \AA\ to 2.496(1) \AA\ ($\Delta R_{\mathrm{OHO}} = 0.033(2)$ \AA), whereas the S$-$O bond lengths ($R^\mathrm{H}_\mathrm{SO}$) are practically unchanged (Table \ref{tab:9}). For \TKDS, the unit cell is practically temperature independent, $R_{\mathrm{ODO}}$ shortens very slightly from 2.532(1) \AA\ to 2.520(2) \AA\ ($\Delta R_{\mathrm{ODO}} = -0.012(3)$ \AA) and the $R^\mathrm{D}_\mathrm{SO}$'s increase significantly, what is in accordance with a change of the spontaneous dielectric polarization.

The thermal parameters for H and D are markedly anisotropic. The diagonal factors $U_{x}(T)$ and $U_{z}(T)$ for coordinates $x$ or $z$ in ($a,b$) planes, parallel or perpendicular to O$\cdots$O, respectively, are given in Table \ref{tab:6}. The corresponding half-widths at half-height (HWHH), $\Delta_{\alpha}(T)$ ($\alpha = x,y,z$), of the gaussian profiles
\begin{equation}\label{eq:1}
P(\alpha,T) = \frac{\exp\left[-\alpha^2/U_{\alpha}(T)\right]}{\sqrt{\pi U_{\alpha}(T)}},
\end{equation}
are given in Table \ref{tab:7}. Because the rather modest thermal parameters for heavy nuclei are virtually identical for both isotopomers, $U_{\alpha}(T)$ and $\Delta_{\alpha}(T)$ are largely representative of vibrational displacements. For \TKHS, $U_{x}^\mathrm{H}(5)$ and $U_{x}^\mathrm{H}(293)$ suggest a markedly anharmonic $\nu$OH mode at a low frequency and/or an unresolved splitting. For \TKDS, $U_{x}^\mathrm{D}(14) \gg U_{x}^\mathrm{H}(12)$ is at variance with the mass effect anticipated for usual single or double-well potentials.

\section{\label{sec:3}The infinite square-well}

\begin{figure}[ht]
\includegraphics[angle=0.,scale=0.6]{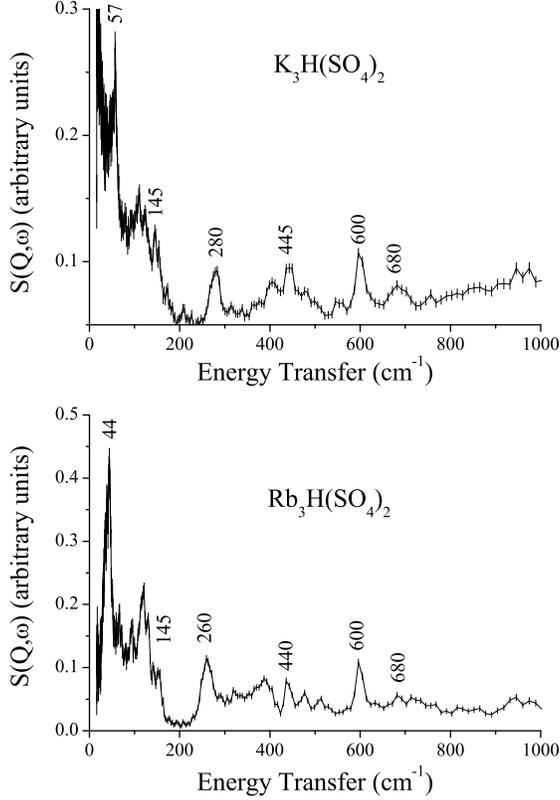}
\caption{\label{fig:TFXA} INS spectra of K$_3$H(SO$_4)_2$ and Rb$_3$H(SO$_4)_2$ crystal powders at 20 K, after ref. [\onlinecite{FLTK}]. The bands at $\approx 600$ \cm\ were assigned to SO$_4$ entities. }
\end{figure}

\begin{figure}[ht]
\includegraphics[angle=0.,scale=0.5]{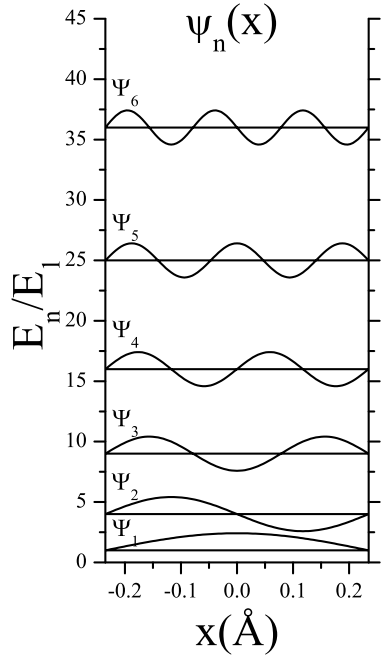}
\caption{\label{fig:PC} Eigenstates of the square-well along the OH stretching coordinate}
\end{figure}

Since the double-well proposed in Ref. [\onlinecite{FLTK}] is rejected by SXND, we seek a single-well potential for a proton oscillator ($m = 1$ amu) consistent with the $\nu$OH transition observed at $(57 \pm 3)$ \cm\ and with the probability density profile. A (quasi)harmonic model is inappropriate because the mean-square amplitude in the ground state, $\langle x^2\rangle_0 \approx 0.29$ \AA$^2$, \footnote{\protect{$\langle u_0^2\rangle = h/\mu\nu \approx 16.715 /\mu\nu$}, $\mu$ in amu and $\nu$ in \cm\ units} should be one order of magnitude greater than $U_{x}^\mathrm{H}(5)$. Consequently, we envisage another textbook potential that is the infinitely deep square-well such that $V = 0$ for $-a_\mathrm{H}/2 \le x \le a_\mathrm{H}/2$ and $V = \infty$ otherwise. \cite{CTDL} The eigenstates are:
\begin{equation}\label{eq:2}\left.\begin{array}{rcl}
E_n & = & \displaystyle{\frac{n^2 h^2}{8ma_\mathrm{H}^2}}\\
\psi_n(x) & = & \displaystyle{\sqrt{\frac{2}{a_\mathrm{H}}}\sin\frac{n\pi (x +a_\mathrm{H}/2)}{a_\mathrm{H}}}\\
\end{array}\right\}; \ n \ge 1.\end{equation}
The zero-point energy is $E_{1}^\mathrm{H} = (E_{2}^\mathrm{H} - E_{1}^\mathrm{H})/3 = (19 \pm 1)$ \cm\ and $a_\mathrm{H} = (0.47 \pm0.01)$ \AA. Table \ref{tab:8} shows that $E_{n}^\mathrm{H} -E_{1}^\mathrm{H} = (n^2 -1)E_{1}^\mathrm{H}$, $2 \le n \leq 6$, compare favorably with INS transitions (Fig. \ref{fig:TFXA}). This is an encouragement to pursue this model. The eigenfunctions are represented in Fig. \ref{fig:PC} and the expected probability density,
\begin{equation}\label{eq:3}
\Psi^2(T) = \frac{\sum_i \exp(-(E_i-E_1)/kT)|\psi_i|^2}{\sum_i \exp(-(E_i-E_1)/kT)},
\end{equation}
can be confronted with (\ref{eq:1}).
\begin{figure}[ht]
\includegraphics[angle=0.,scale=0.5]{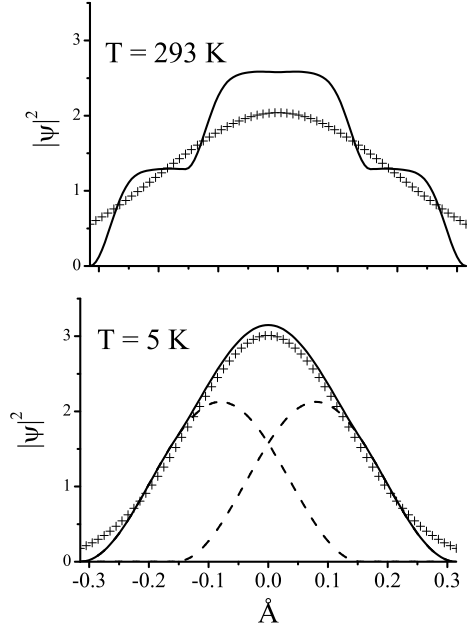}
\caption{\label{fig:TempH} Solid lines: Computed probability densities along the OH stretching coordinate at 5 K (bottom) and 293 K (top) for a rigid H$_{1/2}-$H$_{1/2}$ entity. The length is $l_\mathrm{H} = 0.16$ \AA\ and the center of mass experiences a square-well whose width is $0.47$ \AA. Dashed line: probability density for each half proton. $+$: Gaussian profile whose variance is $U_{x}^\mathrm{H}(T)/2$ (see Table \ref{tab:6}). }
\end{figure}
At 5 K, the leading term $|\psi_1|^2$ gives a bell-shaped profile whose HWHH is $\Delta_{1x}^\mathrm{H} = a_\mathrm{H}/4 \approx 0.12$ \AA. This is significantly less than $\Delta_{x}^\mathrm{H}(5) \approx 0.156$ \AA. We attribute the difference to a splitting of the wavefunction, what means that the energy levels refer to the center of mass of a rigid entity, say H$_{L1/2}-$H$_{R1/2}$, whose separation length is $l_\mathrm{H}$ and whose probability density is the sum of two unresolved profiles $|\psi_1(x \pm l_\mathrm{H}/2)|^2/2$. Our best estimate is $l_\mathrm{H} = (0.16 \pm 0.01)$ \AA\ (see Fig. \ref{fig:TempH}). Then, $\Psi^2(293)$ computed with the same parameters cannot be compared directly to a gaussian profile. The best we can conclude is that the HWHH corresponding to the plateau at half-height, namely $(0.20 \pm 0.05)$ \AA, is in qualitative agreement with $\Delta_{x}^\mathrm{H}(293)$. At this modest level of precision, there is no visible change of $a_\mathrm{H}$ or $l_\mathrm{H}$ consistent with the expansion of the O$\cdots$O bond. There is no visible temperature induced geometric effect.

\begin{figure}[ht]
\includegraphics[angle=0.,scale=0.5]{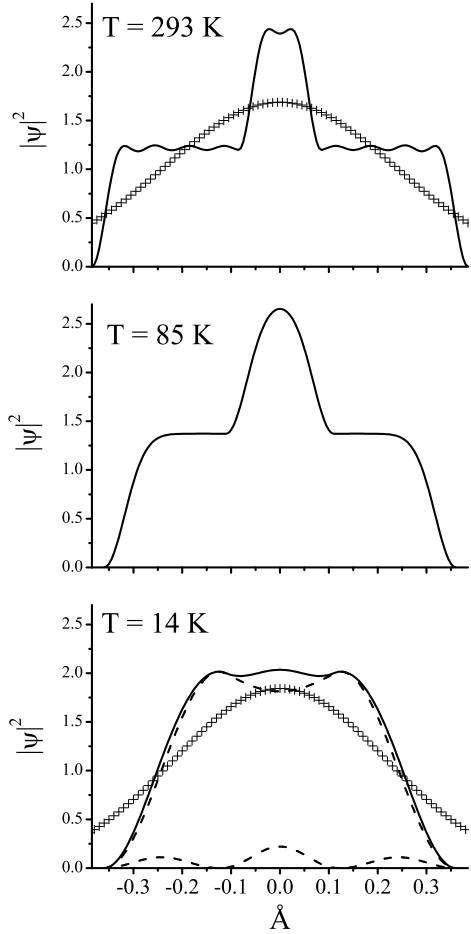}
\caption{\label{fig:TempD}  Solid lines: Computed probability densities along the OD stretching coordinate at 14 K, 85 K, and 293 K for a rigid D$_{L1/2}-$D$_{R1/2}$ entity. The length is $ l_\mathrm{D} = 0.25$ \AA\ and the center of mass experiences a box whose width is $0.47$ \AA. Dashed line: contributions of the $n = 1$ and $n = 2$ states to the probability density. $+$: Gaussian profile whose variance is $U_{x}^\mathrm{D}(T)/2$ (Table \ref{tab:6}). }
\end{figure}

For \TKDS, OD transitions are unknown. We tentatively suppose $a_\mathrm{D} \approx a_\mathrm{H}$, $E_{n}^\mathrm{D} \approx E_{n}^\mathrm{H}/2$, and $|\psi^\mathrm{D}_n|^2 \approx |\psi^\mathrm{H}_n|^2$. Then, $\Delta_{x}^\mathrm{D}(14)$ is consistent with $l_\mathrm{D} \approx 0.25$ \AA\ (Fig. \ref{fig:TempD}). (Note that because of the mass effect, $\Psi^2(14)$ includes a minor contribution of $|\psi_2(x \pm l_\mathrm{D}/2)|^2/2$ to $\approx 5\%$ of the total density). At 293 K, the HWHH of the calculated profile is roughly on the order of $\Delta_{x}^\mathrm{D}(293)$. Note that $l_\mathrm{D} - l_\mathrm{H} \approx 0.09$ \AA\ is rather close to $R_\mathrm{ODO} - R_\mathrm{OHO} \approx 0.07$ \AA\ at low temperatures. This suggests a geometric effect upon isotope substitution.

So far, the analysis of the density profiles is reasonably well established for \TKHS\ at low temperature, since $E_n^\mathrm{H}$ and $l_\mathrm{H}$ are determined by different measurements. For \TKDS\ at low temperature, we ignore whether the assumption $E_n^\mathrm{D} \approx E_n^\mathrm{H}/2$ is correct, so $l_\mathrm{D}$ should be regarded with caution. In addition, it is impossible to conclude whether or not the model is pertinent at elevated temperatures. These drawbacks are wiped away in the next section where we show that dielectric and calorimetric data reported for mixed isotope crystals K$_3$D$_{(1-\rho)}$H$_\rho$(SO$_4)_2$ are precisely represented by this model.

\section{\label{sec:4}Isotope effects and phase transitions}

Because centrosymmetric dimers have no electric dipole, there is no significant coupling terms between proton or deuteron vibrations which are, therefore, degenerate. This is confirmed by INS. The Ising model is, therefore, irrelevant.

Consider an ideal defect-free crystal of \TKDS\ represented by a periodic lattice in 3-dimension of $\mathcal{N}$ indistinguishable rigid entities, D$_{L1/2j}-$D$_{R1/2j}$, centered at every node $j$. $\mathcal{N}$ is on the order of Avogadro's constant. The centrosymmetric wavefunction for the center of mass $x_j$ is
\begin{equation}\label{eq:4}
\psi_{nj}(x_j,l_\mathrm{D}) = \frac {1} {\sqrt{2}} [\psi_{n}(x_j + l_\mathrm{D}/2) - \psi_{n}(x_j - l_\mathrm{D}/2)],
\end{equation}
The lattice can be represented by $\mathcal{N}$ Bloch-wavefunctions indexed $r$ ($1 \leq r \le \mathcal{N}$),
\begin{equation}\label{eq:5}
\Psi_n(x,l_\mathrm{D},\mathbf{k}_{nr}) = \displaystyle{\frac {1} {\mathcal{N}}} \sum\limits_{j=1}^\mathcal{N} \psi_{nj}(x_j,l_\mathrm{D}) \exp i \mathbf{k}_{nr} \mathbf{.L}_j,
\end{equation}
where $x$ is a nonlocal observable independent of $j$, $\mathbf{k}_{nr}$ is a wave vector and $\mathbf{L}_j$ is a lattice vector. Bloch-states are degenerate with respect to $\mathbf{k}_{nr}$, so the interpretation presented below is $\mathbf{k}_{nr}$ independent. For \TKHS, Bloch-states should be antisymmetrized if protons effectively behave as fermion, but this would be of no consequence to our interpretation.

It is worth emphasizing that a symmetric double-well giving rise to tunneling is excluded because there is no symmetry plane perpendicular to O$\cdots$O compatible with a symmetric ground state $2^{-1/2}[|L\rangle + |R\rangle]$. \footnote{The same argument holds for the KDP family} Likewise, symmetric displacements of D$_{L1/2j}$ with respect to D$_{R1/2j}$ are forbidden. Only displacements of the center of mass of the rigid entity are allowed.

\begin{figure}[ht]
\includegraphics[angle=0.,scale=0.3]{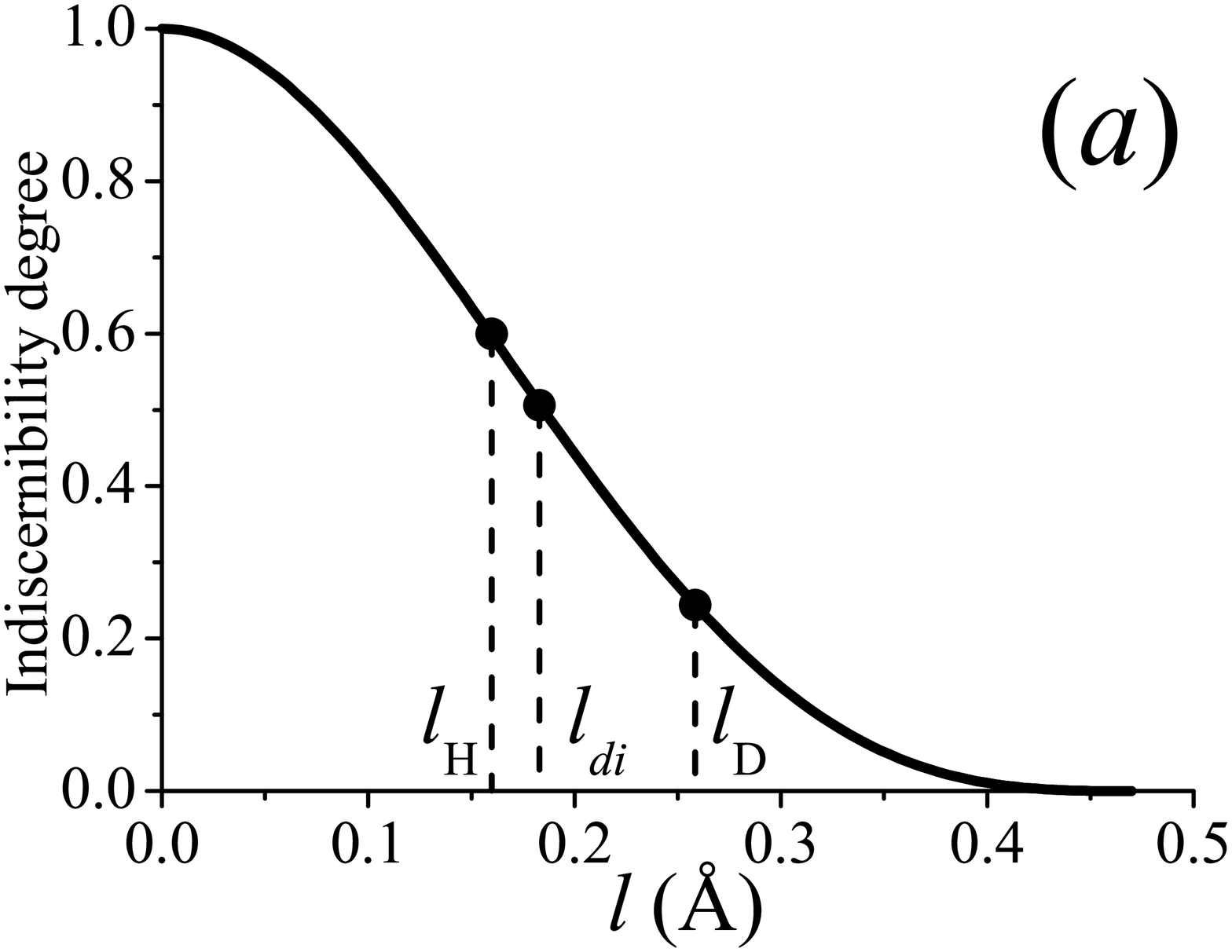}
\includegraphics[angle=0.,scale=0.3]{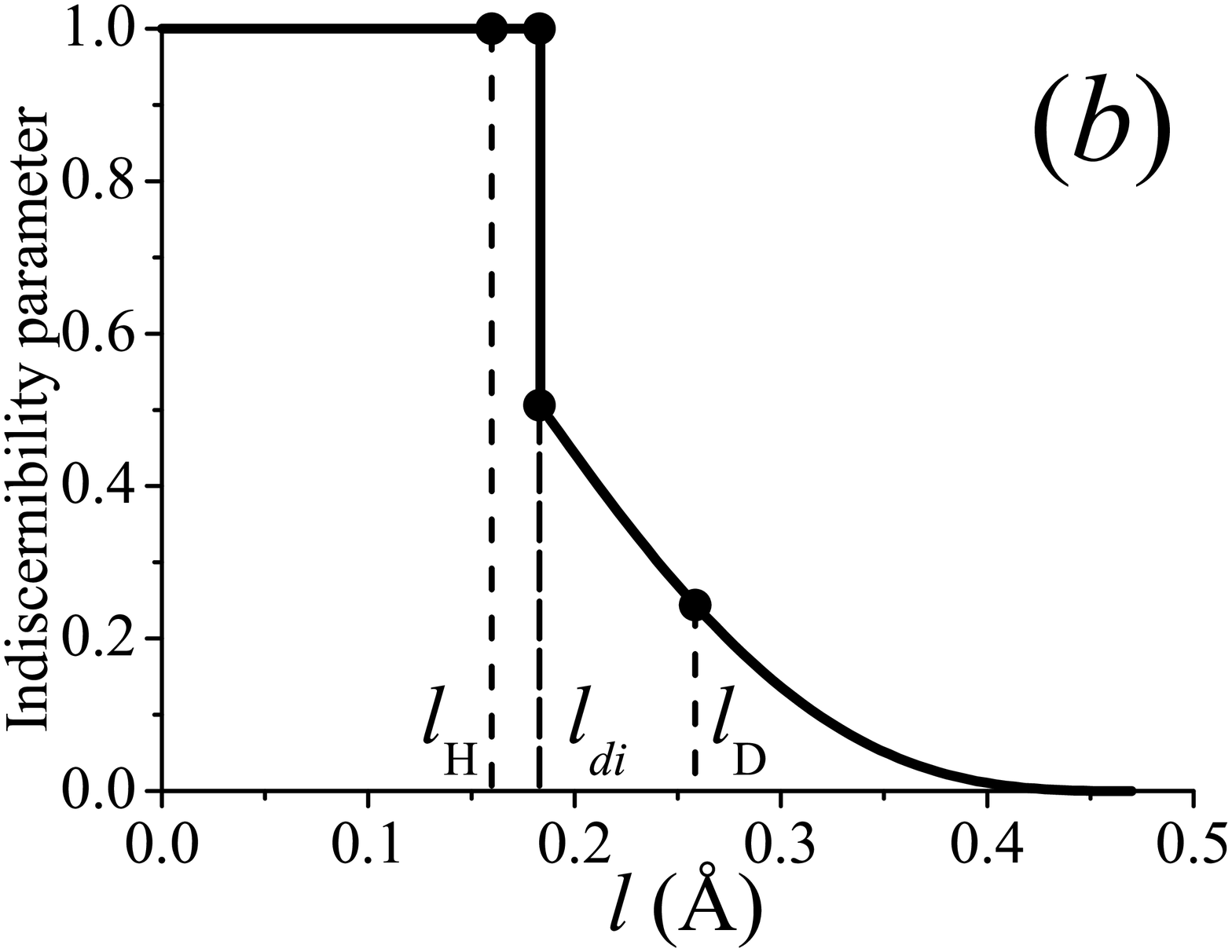}
\caption{\label{fig:I1}  ($a$) The indiscernibility degree of the ground state, $\mathcal{I}_1(l)$, as a function of the separation of the split-wave. $l_\mathrm{H} = 0.16$ \AA, $\mathcal{I}_1 = 0.60$ for K$_3$H(SO$_4)_2$. $l_\mathrm{D} = 0.25$ \AA, $\mathcal{I}_1 = 0.27$ for K$_3$D(SO$_4)_2$. $l_{di} = 0.183$ \AA, $\mathcal{I}_{1di} = 0.50$ at the discernible-indiscernible boundary. ($b$) The transition parameter $\mathcal{IP} = \mathcal{I}_1$ if $l \leq l_{di}$, $\mathcal{IP} = 1$ if $l \geq l_{di}$ . }
\end{figure}

Below $T_c$, \TKHS\ and \TKDS\ demonstrate different probability densities for H or D, different S$-$O bond lengths, opposite temperature effects for their dielectric constants ($d\epsilon^\mathrm{H}/dT < 0$ while $d\epsilon^\mathrm{D}/dT > 0$) and $\epsilon^\mathrm{D}$ reaches a maximum at $T = T_c$. Above $T_c$, probability densities, bond lengths and temperature effects for the dielectric constants are similar: $d\epsilon^\mathrm{H}/dT < 0$; $d\epsilon^\mathrm{D}/dT < 0$. From these observations we tentatively infer that the S$-$O bond lengths, hence the dielectric constants, could correlate with the splitting of the wavefunction or, more specifically, with the nonlocal ``indiscernibility degree'' in the ground state proportional to the overlap defined on a $0-1$ scale as:
\begin{equation}\label{eq:6}\left.\begin{array}{rcl}
\mathcal{I}_1(l) & = & \displaystyle{\frac{2}{a} \int\limits_{-(a-l)/2}^{(a-l)/2} \sin\frac{\pi(x+a/2+l/2)}{a} \sin\frac{\pi(x+a/2-l/2)}{a} dx}\\
& = & \displaystyle{\frac{a-l}{a} \cos\frac{\pi l}{a} +\frac{1}{\pi}\sin \pi \frac{a-l}{a}} \\
\end{array}\right\}. \end{equation}
Then, we tentatively set the discernibility-indiscernibility boundary (DIB) to $\mathcal{I}_1(l_{di}) = 1/2$, for $l_{di} = 0.183$ \AA. Consequently, $\mathcal{I}_1(l_\mathrm{H}) > 1/2$ at every temperature (see Fig. \ref{fig:I1} $a$) and the S$-$O bonds are virtually temperature independent. In contrast, $\mathcal{I}_1(l_\mathrm{D}) < 1/2$ at 14 K and we assume that indiscernibility emerges at elevated temperatures from a thermal pure quantum state leading to coherent oscillations of the center of mass:
\begin{equation}\label{eq:7}\left.\begin{array}{lcr}
|\Psi^\mathrm{D}(t)\rangle & = & \sum\limits_{n \ge 1} \alpha_n^\mathrm{D}|\psi_n \rangle \exp(iE_n^\mathrm{D}t/\hbar)\\
|\alpha_n^\mathrm{D}|^2 & = & \displaystyle{\frac{\exp[-(E_n^\mathrm{D}- E_1^\mathrm{D})/kT]}{\sum\limits_{n\geq 1} \exp[-(E_n^\mathrm{D}- E_1^\mathrm{D})/kT]}}\\
\end{array}\right\}.\end{equation}
The temperature law for the indiscernibility degree is then
\begin{equation}\label{eq:8}
\mathcal{I}^\mathrm{D}(T) = |\alpha_1^\mathrm{D}|^2 \mathcal{I}_1(l_\mathrm{D}) + \sum\limits_{n>1} \sum\limits_{m>n} \beta_{n,m} |\alpha_m^\mathrm{D}|^2,
\end{equation}
where $\beta_{n,m}$ is the indiscernibility degree of the superposition $|n\rangle + e^{i\varphi}|m\rangle$ for which the mean position of the center of mass is: \cite{CTDL}
\begin{equation}\label{eq:9}\left.\begin{array}{lll}
\langle x \rangle_{n,m} (t) & = & \displaystyle{X_{n,m}\cos \left(\frac{E_m^\mathrm{D} - E_n^\mathrm{D}}{\hbar}t + \varphi\right)};\\
X_{n,m} & = & \displaystyle{\frac{2}{a}\int\limits_0^a x \sin\frac{n\pi x}{a} \sin\frac{m\pi x}{a} dx}\\
& = & \displaystyle{\frac{a}{[(n-m)\pi]^2}\left[\cos(n-m)\pi - 1 \right] - \frac{a}{[(n+m)\pi]^2}\left[\cos(n+m)\pi - 1 \right] }\\
\end{array}\right\}.\end{equation}
If $m \neq n+1$, $X_{n,m} \approx 0$ and $\beta_{n,m} \approx 0$. Otherwise, $X_{n,n+1} \approx -2a\pi^{-2}$ and the mean amplitude of oscillations, $\langle l\rangle = 2\sqrt{2}a\pi^{-2}$, determines the time-averaged mean-separation of the split-wave: $\langle l_\mathrm{D}\rangle = l_\mathrm{D} - \langle l\rangle = (0.11 \pm 0.01)$ \AA. Then, $\beta_{n,n+1} = \mathcal{I}_1(\langle l_\mathrm{D}\rangle) = 0.77 \pm 0.03$. The DIB $\mathcal{I}^\mathrm{D}(T_{di}) = 1/2$ gives $T_{di} = (86 \pm 10)$ K, in accordance with $T_c$. Our assumptions $\mathcal{I}_1(l_{di}) = 1/2$ and $E_n^D \approx E_n^H/2$, our estimate $l_\mathrm{D} = 0.25$ \AA\ and the thermal pure quantum state (\ref{eq:7}), are thus validated at the macroscopic level of the phase transition. We conclude that the square-well is largely isotope and temperature independent, to the least below $T_{di}$.

Our interpretation of the isotope effect is as follows. For \TKHS, $\mathcal{I}_1 (l_\mathrm{D}) > 1/2$ at every temperature, the S$-$O bonds are practically temperature independent, and the monotonous decrease of $\epsilon^\mathrm{H}$ upon rising $T$ is due to thermal fluctuations of the crystal lattice. For \TKDS, $\mathcal{I}_1 (l_\mathrm{D}) < 1/2$ for $T < T_{di}$, the S$-$O bonds are longer and $\epsilon^\mathrm{D} < \epsilon^\mathrm{H}$. As $T$ increases, $\mathcal{I}^\mathrm{D}(T)$ increases, the S$-$O bonds shorten, $\epsilon^\mathrm{D}$ increases and reaches a maximum at $T_{di} = T_c$. Since the probability density computed at 85 K is similar to that at 293 K (see Fig. \ref{fig:TempD}), we infer that the bond lengths are practically temperature independent above $T_{di}$, so the monotonous decrease of $\epsilon^\mathrm{D}$ is analogous to that of $\epsilon^\mathrm{H}$.

\begin{figure}[ht]
\includegraphics[angle=0.,scale=0.3]{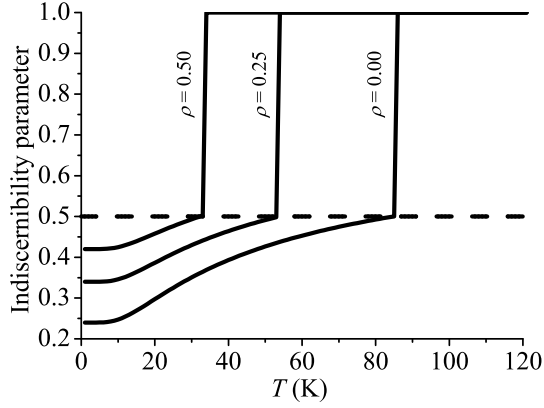}
\caption{\label{fig:Ind}  Temperature laws for the indiscernibility parameter driving the phase transition of K$_3$D$_{(1-\rho)}$H$_\rho$(SO$_4)_2$, according to eqs (\ref{eq:8}) and (\ref{eq:10}). Dash: the dividing line between discernibility (below) and indiscernibility (above) domains. }
\end{figure}

So far, the cusp at $T = T_{di}$ for $\epsilon^\mathrm{D}$ reveals a discontinuity that is not accounted for by (\ref{eq:8}). Consequently, we suppose that the phase transition is actually triggered by a discontinuity of the indiscernibility parameter, say $\mathcal{IP}_1$, such that: (i) $\mathcal{IP}_1 = \mathcal{I}_1$ if $l \geq l_{di}$ or $\mathcal{IP}_1 = 1$ if $l \leq l_{di}$ (see Fig. \ref{fig:I1} $b$); (ii) $\mathcal{IP}^\mathrm{D}(T) = \mathcal{I}^\mathrm{D}(T)$ and $\beta_{n,n+1} = \mathcal{I}_1(\langle l_\mathrm{D}\rangle)$, if $T \leq T_{di}$, or $\mathcal{IP}^\mathrm{D}(T) = 1$ and $\beta_{n,n+1} = 1$ if $T \geq T_{di}$ (see Fig. \ref{fig:Ind}). This means that indiscernibility is quantified in the discernible domain whereas it is complete in the indiscernible range. This interpretation is corroborated by calorimetric measurements. For $T < T_{di}$, we can label 1 or 2, respectively, the components of the split wavefunction (\ref{eq:4}) and $\psi_n(x_{j1}-l_\mathrm{D}/2)- \psi_n(x_{j2}+l_\mathrm{D}/2) \neq \psi_n(x_{j2}-l_\mathrm{D}/2) - \psi_n(x_{j1}+l_\mathrm{D}/2)$. Otherwise, for $T > T_{di}$, $\psi_n(x_{j1}-l_\mathrm{D}/2)- \psi_n(x_{j2}+l_\mathrm{D}/2) \equiv \psi_n(x_{j2}-l_\mathrm{D}/2) - \psi_n(x_{j1}+l_\mathrm{D}/2)$. Consequently, the total molar-entropy of the transition is $R\ln2$ and the specific heat jump is $\Delta C = 3R/2$, as effectively measured. \cite{MTNSK}

\begin{figure}[ht]
\includegraphics[angle=0.,scale=0.3]{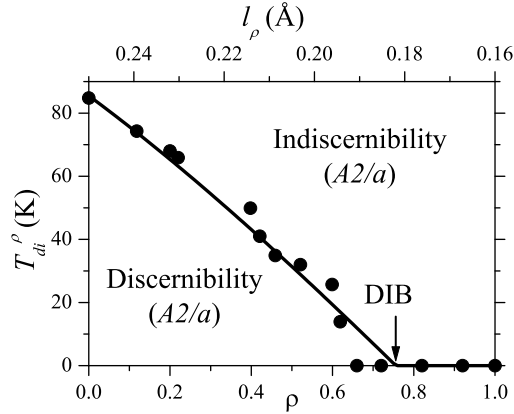}
\caption{\label{fig:Tc}  Solid line: Temperature of the discernible-indiscernible transition for K$_3$D$_{(1-\rho)}$H$_\rho$(SO$_4)_2$, according to eq. (\ref{eq:11}). $\bullet$: experimental data (dielectric measurements) estimated from digitized Fig. 3 in Ref. [\onlinecite{MTNSK}]. DIB: discernible-indiscernible boundary. $l_\rho = (1-\rho)l_\mathrm{D} + \rho l_\mathrm{H}$ is the effective separation length for the components of the wavefunction.}
\end{figure}

The indiscernibility parameter is also relevant for K$_3$D$_{(1-\rho)}$H$_\rho$(SO$_4)_2$. Dielectric measurements realize Bloch-states at the center of the Brillouin-zone, via insignificant energy or momentum transfer, and variables of statistical-mechanical interest can be calculated from only one realization of the thermal pure quantum state. \cite{SS} Protons and deuterons are not separate, so the indiscernibility parameter at 0 K is $\mathcal{IP}_1(l_\rho)$, where $l_\rho = \rho l_\mathrm{H} + (1-\rho)l_\mathrm{D}$. The DIB corresponds to a quantum phase transition at 0 K for $\rho_{di} = 0.74 \pm 0.02$. The temperature law analogous to (\ref{eq:8}),
\begin{equation}\label{eq:10}\left.\begin{array}{ll}
\left.\begin{array}{lll}\mathcal{IP}^\rho(T < T_{di}^\rho) & = & |\alpha_1^\mathrm{D}|^2 \mathcal{IP}_1(l_\rho) + \sum\limits_{n>1} \beta |\alpha_n^\mathrm{D}|^2 \\
\mathcal{IP}^\rho(T > T_{di}^\rho) & = & 1 \\
\end{array}\right\}\ & 0 \leq \rho < \rho_{di}\\
\mathcal{IP}^\rho(T) = 1 & \rho_{di} < \rho \leq 1\\
\end{array}\right\},\end{equation}
accounts for the thermally induced phase transition at $T_{di}^\rho$ for $\rho < \rho_{di}$ and for the lack of transition for $\rho > \rho_{di}$. Here, we  assume that $E_n^\mathrm{H}$ and $E_n^\mathrm{D}$ realized through energy transfer are separable and independent of $\rho$. Visual examination of Fig. \ref{fig:Ind} shows that for $T < T^\rho_{di}$, $\mathcal{IP}^\rho$ is shifted upward by $\mathcal{IP}_1 (l_\rho)- \mathcal{IP}_1(l_0)$ ($l_0 = l_\mathrm{D}$) and $T_{di}^\rho$ can be written as:
\begin{equation}\label{eq:11}\left. \begin{array}{rclr}
T_{di}^\rho & \approx & \displaystyle{\frac{1/2 - \mathcal{IP}_1(l_\rho)}{1/2 -\mathcal{IP}_1(l_0)} T_{di}^0};  & 0\leq \rho < \rho_{di}\\
T_{di}^\rho & = & 0; & \rho_{di} < \rho \leq 1\\
\end{array}\right\}.\end{equation}
The quasi-linear variation of $T_{di}^\rho$ for $\rho < \rho_{di}$, mirroring that of $\mathcal{IP}_1(l_\rho)$, is in reasonably good agreement with measurements (see Fig. \ref{fig:Tc}). In contrast, the Ising model advocated by Moritomo et al. yields a curved function with a very steep slope as $T_{c}(\rho) \longrightarrow 0$ (see Fig. 3 in Ref. [\onlinecite{MTNSK}]), so the extrapolated critical concentration ($\rho_c = 0.66\pm0.04$) \cite{MTNSK} is smaller than $\rho_{di}$. In practice, dielectric data suggest that it is technically difficult to measure $T_{di} < \approx 10$ K with precision, so $\rho_c$ is logically a lower bound for $\rho_{di}$.

\begin{figure}[ht]
\includegraphics[angle=0.,scale=0.3]{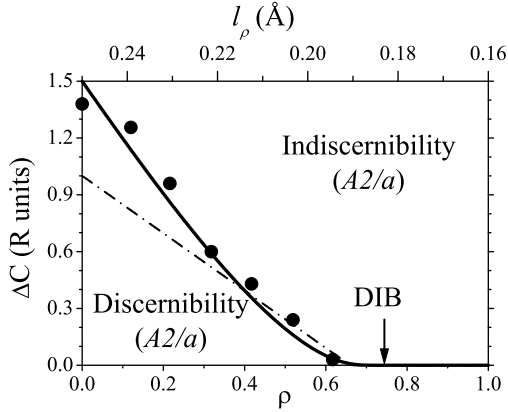}
\caption{\label{fig:C}  Solid: specific heat jump for K$_3$D$_{(1-\rho)}$H$_\rho$(SO$_4)_2$, according to eq. (\ref{eq:13}). Dot dash: the Ising model, as shown in Ref. [\onlinecite{MTNSK}]. $\bullet$: experimental data estimated from digitized Fig. 4 in Ref. [\onlinecite{MTNSK}]. DIB: discernible-indiscernible boundary. $l_\rho = (1-\rho)l_\mathrm{D} + \rho l_\mathrm{H}$ is the effective separation length for the components of the wavefunction.}
\end{figure}

According to Fig. \ref{fig:Ind}, the specific heat jump $\Delta C (\rho)$ for $\rho < \rho_{di}$ is proportional to
\begin{equation}\label{eq:12}
\int\limits_0^{T_{di}^\rho} \mathcal{IP}^\rho(T)dT \approx T_{di}^\rho [1/2 - \mathcal{IP}_1(l_\rho)],
\end{equation}
so
\begin{equation}\label{eq:13}\left.\begin{array}{llll}
\Delta C(\rho) & \approx & \displaystyle{\frac{3R}{2} \times \frac{T_{di}^\rho [1/2 - \mathcal{IP}_1(l_\rho)]} {T_{di}^0 [1/2 - \mathcal{IP}_1(l_0)]}};\ & 0\leq \rho < \rho_{di}\\
\Delta C(\rho) & = & 0; & \rho_{di} < \rho \leq 1\\
\end{array}\right\}.\end{equation}
This is in reasonably good agreement with observations (see Fig. \ref{fig:C}). The shape of $\Delta R(\rho)$ mirrors that of $\mathcal{IP}^0(T)$ in Fig. \ref{fig:Ind}. It is nearly linear for $\rho < \approx 0.4$ and the slope goes to zero as $\mathcal{IP}_1(l_\rho) \longrightarrow \mathcal{IP}_1(l_{di})$, so there is no discontinuity at the DIB.

\section{\label{sec:5}R$\mathbf{b}_3$H(SO$_4)_2$}

The INS spectrum of the isomorphous crystal of Rb$_3$H(SO$_4)_2$ (see Fig. \ref{fig:TFXA} and Table \ref{tab:6}) is also consistent with a square-well, apart from the transition observed at $\approx 44$ \cm, instead of $57 $ \cm. However, the high background casts doubt on the actual frequency that should be reassessed with a more appropriate instrument. \cite{FLTK} In any case, an OH band in such a low frequency range is in conflict with the interpretation of NCS data proposed by Homouz et al. \cite{HREMB} for a single-crystal of \TRbHS\ at 10 K. These authors observed that the mean proton-momentum values along the crystal axes are rather similar ($\sigma_{b} = 4.60$ \A, $\sigma _{a} = 3.73$ \A, $\sigma_{c^*} = 4.35$ \A), from what they inferred that the three OH modes should be in the 1000-1250 \cm\ range. This is definitively a logical fallacy because the OH modes are not parallel to the crystal axes. In fact, we can estimate the mean-momentum for each OH mode from the known frequencies. With the square-well model, $\langle p_x^2\rangle_1^{1/2} = h/2a \approx 1$ \A. \cite{CTDL} For the bending modes observed at $\approx 1250$ \cm\ and $\approx 1600$ \cm\ for both \TKHS\ and \TRbHS, the harmonic approximation gives $\langle p^2_y \rangle_0^{1/2} \approx 4.39$ \A\ and $\langle p^2_z \rangle_0^{1/2} \approx 5.00$ \A. Then, the projections along the crystal axes are:

\begin{equation}\label{eq:14}\left . \begin{array}{lcll}
\langle p_b^2 \rangle^{1/2} & = & \langle p_x^2 \rangle_1^{1/2} \cos30^\circ + \langle p_z ^2 \rangle_0^{1/2} \sin30^\circ & \approx 3.37\ \mathrm{\AA}^{-1}\\
\langle p_a^2 \rangle^{1/2} & = & \langle p_x^2 \rangle_1^{1/2} \sin30^\circ + \langle p_z ^2 \rangle_0^{1/2} \cos30^\circ & \approx 4.83\ \mathrm{\AA}^{-1}\\
\langle p_{c^*}^2 \rangle^{1/2} & = & \langle p_y^2 \rangle_0^{1/2} & \approx 4.39\ \mathrm{\AA}^{-1}\\
\end{array}\right\}.\end{equation}
These projections are effectively rather similar although the distribution is markedly anisotropic. Furthermore, the low $\nu$OH frequency imposes $\sigma_b < \sigma_a$, so we suspect $\sigma_a$ and $\sigma_b$ were swaped by Homouz et al. Then, taking into account this permutation, numerics (\ref{eq:14}) are in reasonably good agreement with NCS values: $\sigma_a-\langle p_b^2 \rangle_1^{1/2} = 0.36$ \A; $\sigma_b-\langle p_a^2 \rangle_1^{1/2} = -0.23$ \A; $\sigma_x-\langle p_{c^*}^2 \rangle_1^{1/2} = -0.04$ \A. The conflict of interpretation put forward by Homouz et al. is pointless and there is every reason to suppose that the square-well holds for \TRbHS. Our interpretation should be relevant for the dielectric anomaly of \TRbDS\ observed at 82 K. 

\section{\label{sec:6}Conclusion}

The space group symmetry of K$_3$H(SO$_4)_2$ or K$_3$D(SO$_4)_2$ determined with neutron diffraction is $A2/a$ at cryogenic or room temperature and there is no resolvable site splitting for H or D. Statistical disorder and antiferroelectricity are ruled out. The anomalies of the dielectric constants or heat capacity do not correspond to any structural symmetry breaking.

At low temperatures, the probability densities for H or D and the INS spectrum of K$_3$H(SO$_4)_2$ are consistent with split-waves representing rigid entities H$_{L1/2} - $H$_{R1/2} $ ($l_\mathrm{H} \approx 0.16$ \AA) or D$_{L1/2} - $D$_{R1/2} $ ($l_\mathrm{D} \approx 0.25$ \AA) whose center of mass experiences a square-well. We leave open the question as to whether computational chemistry could confirm or not this counterintuitive well, as opposed to double-wells for longer bonds or funnel-shaped single-wells for shorter bonds.

Our explanation of the dielectric and calorimetric anomalies is based upon two postulates. The first one is that these anomalies correspond to the crossing of a definite boundary for the discernibility-indiscernibility degree. To the best of our knowledge, this is unprecedented and we ignore whether or not this boundary is deeply rooted in the ground of quantum mechanics. The indiscernibility parameter can be thought of as a quantum counterpart to an order parameter diverging at the phase transition in classical physics. The second postulate is the correlation of the indiscernibility parameter and the dielectric constant, via the S$-$O bonds. The physical ground of this correlation is largely unknown. Our theory is scaled by three measured parameters, $a$, $l_\mathrm{H}$, $l_\mathrm{D}$. It accounts precisely for every observation, within the framework of the quantum theory of measurements. The crystal is a macroscopic-scale quantum system for which the existence of a thermal pure quantum state is allowed by the adiabatic separation of H or D and heavy nuclei, as commonly encountered in similar O$\cdots$O bonds.

\begin{table}[p]
\caption{\label{tab:1} Neutron single crystal diffraction data and structure refinement for K$_3$H(SO$_4)_2$ and K$_3$D(SO$_4)_2$.
 $\lambda$ = 0.8305 \AA. Space groups monoclinic $A 2/a$ with Z = 8. The criterion for used reflections was I $>$ 3$\sigma $(I).
  The variance for the last digit is given in parentheses. Refinement on F.}

\begin{center}
\begin{ruledtabular}
\begin{tabular}{llllllllll}
\hline
& \multicolumn{4}{l}{K$_3$H(SO$_4)_2$} & \ & \multicolumn{3}{l}{K$_3$D(SO$_4)_2$ } \\
 \cline{2-6} \cline{7-10}
 & 5 K & \ \ & 12 K & \ \ & 293 K & \ & 14 K & \ & 293 K\\
\cline{2-2} \cline{4-4} \cline{6-6} \cline{8-8} \cline{10-10}
a ({\AA}) & 9.704(1) & & 9.700(1) & & 9.806(1) & & 9.777(1) & & 9.778(1) \\
b ({\AA}) & 5.639(1) & & 5.634(1) & & 5.687(1) & & 5.674(1) & & 5.681(1) \\
c ({\AA}) & 14.526(1) & & 14.524(1)  & & 14.702(1) & & 14.667(1) & & 14.701(1) \\
\textit{$\beta $ }(\r{ }) & 102.95(1) & & 102.73(1)  & & 102.93(1) & & 102.97(1) & & 103.05(1) \\
$V$ ({\AA}$^{3})$ & 774.5(2) & & 774.3(2) & & 799.1(2) & & 792.9(2) & & 795.5(2) \\
%$D_{x}$ & 2.66 & & 2.66  & & 2.58 & & 2.61 & & 2.60 \\
 & & & & & & & \\
Measured reflections & 2100 & & 2385  & & 2687 & & 1822 & & 794 \\
Independent reflections & 1749 & & 1373  & & 1373 & & 1253 & & 432 \\
Used reflections & 1524 & & 1183  & & 1781 & & 999 & & 404 \\
Coefficients & 61 & & 61 & & 61 & & 61 & \ & 61 \\
& & & & & & & \\
$\mathrm{R_{int}}$ & 0.0002 & & 0.0015  & & 0.0002 & & 0.0004 & & 0.0006 \\
R-factor & 0.027 & & 0.045  & & 0.024 & & 0.047 & & 0.018 \\
Weighted R-factor & 0.023 & & 0.051  & & 0.014 & & 0.033 & & 0.020 \\
Goodness of fit & 1.101 & & 0.885  & & 1.077 & & 1.066 & & 1.064 \\
Extinction coefficient & 13.8(5) & & 14.8(10)  & & 9.6(4) & & 5.0(3) & & 40(1) \\

\end{tabular}
\end{ruledtabular}
\end{center}
\end{table}

\begin{table}[p]
\caption{\label{tab:2} Atomic positions and isotropic
temperature factors for K$_3$H(SO$_4)_2$ at 5 K
(first lines), 12 K (second lines) and 293 K (third lines).}

\begin{center}
\begin{ruledtabular}
\begin{tabular}{lllll}

 Atom & $x/a$ & $y/b$ & $z/c$ & U(iso)(\AA$^2$) \\

\hline

K(1) & 0.2500     & 0.75077(11) & 0.0000 & 0.0041 \\
     & 0.2500 & 0.7508(2) & 0.0000 & 0.0048 \\
     & 0.2500 & 0.75551(13) & 0.0000 & 0.0198 \\
K(2) & 0.34809(5) & 0.73247(8) & 0.30601(3) & 0.0044 \\
     & 0.34802(9) & 0.73254(16) & 0.30583(7) & 0.0047 \\
     & 0.34786(5) & 0.73208(8) & 0.30578(4) & 0.0226 \\
S(1) & 0.03807(6) & 0.2726(1) & -0.11626(4) & 0.0025 \\
     & 0.03802(11) & 0.2725(2) & -0.11641(8) & 0.0024 \\
     & 0.03807(5) & 0.27302(8) & -0.11599(3) & 0.0131 \\
O(1) & 0.09896(3) & 0.47808(5) & -0.15424(2) & 0.0050 \\
     & 0.09897(6) & 0.4780(1) & -0.15423(5) & 0.0055 \\
     & 0.09819(3) & 0.47694(5) & -0.15331(2) & 0.0205 \\
O(2) & 0.10868(3) & 0.05211(5) & -0.13025(2) & 0.0051 \\
     & 0.10875(5) & 0.0521(1) & -0.13021(4) & 0.0053 \\
     & 0.10805(4) & 0.05494(5) & -0.13010(2) & 0.0224 \\
O(3) & -0.11553(3) & 0.25992(5) & -0.15300(2) & 0.0050 \\
     & -0.11559(6) & 0.2599(1) & -0.15298(4) & 0.0052 \\
     & -0.11381(3) & 0.26055(6) & -0.15244(2) & 0.0205 \\
O(4) & 0.06148(3) & 0.31314(5) & -0.009127(19) & 0.0052 \\
     & 0.06141(6) & 0.31321(11) & -0.00912(4) & 0.0054 \\
     & 0.06099(4) & 0.31245(7) & -0.010210(19) & 0.0241 \\
H(1) & 0.0000 & 0.5000 & 0.0000 & 0.0213 \\
     & 0.0000 & 0.5000 & 0.0000 & 0.0222 \\
     & 0.0000 & 0.5000 & 0.0000 & 0.0477 \\

\end{tabular}
\end{ruledtabular}
\end{center}
\end{table}

\begin{table}[p]
\caption{\label{tab:3} Atomic positions and isotropic
temperature factors for K$_3$D(SO$_4)_2$ at 14 K
(first lines) 293 K (second lines).}

\begin{center}
\begin{ruledtabular}
\begin{tabular}{lllll}

 Atom & $x/a$ & $y/b$ & $z/c$ & U(iso)(\AA$^2$) \\

\hline

K(1) & 0.2500 & 0.7517(3) & 0.0000 & 0.0042 \\
     & 0.2500 & 0.7559(2) & 0.0000 & 0.0217\\
K(2) & 0.34821(12) & 0.7343(2) & 0.30569(9) & 0.0052 \\
     & 0.34804(11) & 0.73270(17) & 0.30574(8) & 0.0242 \\
S(1) & 0.03846(15) & 0.2704(3) & -0.1164(1) & 0.0027 \\
     & 0.03823(13) & 0.2727(2) & -0.11604(8) & 0.0142 \\
O(1) & 0.09942(8) & 0.47698(14) & -0.15382(5) & 0.0048 \\
     & 0.09827(6) & 0.47659(11) & -0.15307(4) & 0.0219 \\
O(2) & 0.10852(9) & 0.05043(15) & -0.13138(5) & 0.0057 \\
     & 0.10803(6) & 0.05459(12) & -0.13039(5) & 0.0238 \\
O(3) & -0.11496(8) & 0.25884(15) & -0.15246(5) & 0.0050 \\
     & -0.11348(7) & 0.26004(11) & -0.15222(4) & 0.0221 \\
O(4) & 0.06275(8) & 0.30902(15) & -0.00884(5) & 0.0055 \\
     & 0.06161(7) & 0.31036(16) & -0.01022(4) & 0.0261 \\
D(1) & 0.0000 & 0.5000 & 0.0000 & 0.0396 \\
     & 0.0000 & 0.5000 & 0.0000 & 0.0533 \\

\end{tabular}
\end{ruledtabular}
\end{center}
\end{table}

\begin{table}[p]
\caption{\label{tab:4} Thermal parameters in \AA$^2$ units for K$_3$H(SO$_4)_2$ at 5 K (first lines), 12  K (second lines) and 293 K (third lines). The thermal parameters account for the variation of the contribution of each atom to Bragg's peak intensities through the thermal factor $T^{at}$ depending on the reciprocal lattice parameters $a^*$, $b^*$, $c^*$, and unit cell indexes in reciprocal space $h, k, l$, as $T^{at} = \exp [ -2\pi^2 (U_{11}^{at}h^2a^{*2} + U_{22}^{at}k^2b^{*2} + U_{33}^{at}l^2c^{*2}  + 2U_{12}^{at}ha^*kb^* + 2U_{23}^{at}kb^*lc^* + U_{31}^{at}lc^*ha^* ) ]$.}

\begin{center}
\begin{ruledtabular}
\begin{tabular}{lllllll}

 Atom & U$_{11}$ & U$_{22}$ & U$_{33}$ & U$_{23}$ & U$_{13}$ & U$_{12}$ \\

\hline

K(1) & 0.0043(2) & 0.0048(2) & 0.0031(2) & 0.0000 & 0.00068(16) & 0.0000 \\
     & 0.0037(5) & 0.0046(5) & 0.0062(6) & 0.0000 & 0.0010(4) & 0.0000 \\
     & 0.0184(2) & 0.0242(2) & 0.0165(2) & 0.0000 & 0.00314(16) & 0.0000 \\
K(2) & 0.00446(16) & 0.00402(16) & 0.00458(15) & -0.00050(12) & 0.00104(12) & -0.00024(12) \\
     & 0.0042(4) & 0.0037(4) & 0.0063(4) & -0.0008(3) & 0.0010(3) & -0.0004(2) \\
     & 0.01899(15) & 0.02015(16) & 0.02811(19) & -0.00470(15) & 0.00395(14) & -0.00035(14) \\
S(1) & 0.00272(19) & 0.00217(18) & 0.00252(18) & 0.00001(14) & 0.00057(14) & 0.00014(15) \\
     & 0.0031(4) & 0.0023(5) & 0.0021(5) & -0.0004(3) & 0.0009(3) & 0.0001(3) \\
     & 0.01257(15) & 0.01287(15) & 0.01369(16) & 0.00044(12) & 0.00284(12) & 0.00021(13) \\
O(1) & 0.00599(12) & 0.00410(11) & 0.00525(11) & 0.00060(8) & 0.00187(8) & -0.00108(8) \\
     & 0.0058(2) & 0.0036(3) & 0.0073(3) & 0.00072(17) & 0.00221(18) & -0.00122(15) \\
     & 0.02433(13) & 0.0168(1) & 0.02191(11) & 0.00121(8) & 0.00822(9) & -0.00460(8) \\
O(2) & 0.00578(11) & 0.00396(11) & 0.00583(11) & -0.00002(8) & 0.00169(8) & 0.00145(8) \\
     & 0.0054(2) & 0.0033(3) & 0.0074(3) & 0.00026(17) & 0.00179(19) & 0.00153(15) \\
     & 0.02339(13) & 0.01679(11) & 0.02721(13) & 0.00138(9) & 0.00609(9) & 0.00638(9) \\
O(3) & 0.00338(11) & 0.00607(11) & 0.0053(1) & -0.00023(8) & 0.00057(7) & -0.00014(8) \\
     & 0.0034(2) & 0.0056(3) & 0.0063(3) & -0.00033(17) & 0.00058(17) & -0.00018(14) \\
     & 0.01303(9) & 0.02581(12) & 0.02203(11) & -0.0023(1) & 0.00271(7) & -0.00040(9) \\
O(4) & 0.00666(12) & 0.00599(12) & 0.0029(1) & -0.00011(7) & 0.00090(8) & 0.00021(8) \\
     & 0.0065(2) & 0.0056(3) & 0.0040(3) & -0.00022(17) & 0.00078(17) & 0.00022(15) \\
     & 0.02840(14) & 0.03032(14) & 0.0127(1) & 0.00036(9) & 0.00304(9) & 0.0003(1) \\
H(1) & 0.0209 & 0.0304 & 0.0129 & -0.0019 & 0.0041 & -0.0082 \\
     & 0.0204 & 0.0282 & 0.0178 & -0.0017 & 0.0041 & -0.0073 \\
     & 0.0457 & 0.0765 & 0.0212 & -0.0045 & 0.0081 & -0.0282 \\
\end{tabular}
\end{ruledtabular}
\end{center}
\end{table}

\begin{table}[p]
\caption{\label{tab:5} Thermal parameters in \AA$^2$ units for K$_3$D(SO$_4)_2$ at 14 K (first lines) and 293 K (second lines). The thermal parameters account for the variation of the contribution of each atom to Bragg's peak intensities through the thermal factor $T^{at}$ depending on the reciprocal lattice parameters $a^*$, $b^*$, $c^*$, and unit cell indexes in reciprocal space $h, k, l$, as $T^{at} = \exp [ -2\pi^2 (U_{11}^{at}h^2a^{*2} + U_{22}^{at}k^2b^{*2} + U_{33}^{at}l^2c^{*2}  + 2U_{12}^{at}ha^*kb^* + 2U_{23}^{at}kb^*lc^* + U_{31}^{at}lc^*ha^* ) ]$.}

\begin{center}
\begin{ruledtabular}
\begin{tabular}{lllllll}

 Atom & U$_{11}$ & U$_{22}$ & U$_{33}$ & U$_{23}$ & U$_{13}$ & U$_{12}$ \\

\hline

K(1) & 0.0037(6) & 0.0056(7) & 0.0032(6) & 0.0000 & 0.0006(5) & 0.0000 \\
     & 0.0170(9) & 0.028(1) & 0.0196(8) & 0.0000 & 0.0030(7) & 0.0000 \\
K(2) & 0.0043(4) & 0.0041(5) & 0.0071(4) & -0.0015(4) & 0.0012(3) & -0.0003(4) \\
     & 0.0174(7) & 0.0240(7) & 0.0301(7) & -0.0057(5) & 0.0033(5) & -0.0006(4) \\
S(1) & 0.0023(5) & 0.0027(6) & 0.0026(5) & 0.0008(5) & -0.0002(4) & -0.0004(5) \\
     & 0.0109(7) & 0.0155(7) & 0.0161(7) & 0.0019(5) & 0.0026(5) & 0.0004(4) \\
O(1) & 0.0057(3) & 0.0030(3) & 0.0060(3) & 0.0006(2) & 0.0020(2) & -0.0012(2) \\
     & 0.0227(4) & 0.0200(5) & 0.0243(5) & 0.0010(3) & 0.0079(3) & -0.0051(3) \\
O(2) & 0.0063(3) & 0.0040(3) & 0.0070(3) & 0.0002(3) & 0.0020(2) & 0.0016(3) \\
     & 0.0216(4) & 0.0205(5) & 0.0293(4) & 0.0011(3) & 0.0059(3) & 0.0072(3) \\
O(3) &0.0033(3) & 0.0058(3) & 0.0059(3) & -0.0012(3) & 0.0009(2) & -0.0004(3) \\
     & 0.0115(4) & 0.0298(5) & 0.0243(5) & -0.0028(3) & 0.0029(3) & -0.0009(2) \\
O(4) & 0.0075(3) & 0.0055(3) & 0.0036(3) & -0.0013(3) & 0.0016(2) & 0.0018(3) \\
     & 0.0280(5) & 0.0335(5) & 0.0158(5) & 0.0003(3) & 0.0028(3) & 0.0003(4) \\
D(1) & 0.035(1) & 0.0728(16) & 0.0112(6) & -0.0053(7) & 0.0066(5) & -0.035(1) \\
     & 0.0464(7) & 0.0907(11) & 0.0222(6) & -0.0052(6) & 0.0066(5) & -0.0370(8) \\
\end{tabular}
\end{ruledtabular}
\end{center}
\end{table}

\begin{table}[ht]
\caption{\label{tab:9} Bond lengths in \AA\ units of the SO$_4$ entities and their expansion $\Delta$ with temperature. }
\begin{center}
\begin{ruledtabular}
\begin{tabular}{lllllll}

         & K$_3$H(SO$_4)_2$ & & & K$_3$D(SO$_4)_2$ & & \\
\cline {2-4} \cline {5-7}
         & 5 K & 293 K & $\Delta R^\mathrm{H}_\mathrm{SO}$ & 14 K & 293 K & $\Delta R^\mathrm{D}_\mathrm{SO}$ \\
\cline {2-2} \cline {3-3} \cline {4-4} \cline {5-5} \cline {6-6} \cline {7-7}
S1$-$O1 & 1.4636(6) & 1.4623(5) & 0.0013(11) & 1.4754(17) & 1.4585(14) & 0.0169(31) \\
S1$-$O2 & 1.4555(6) & 1.4545(6) & 0.0010(12) & 1.4637(17) & 1.4527(13) & 0.0110(30) \\
S1$-$O3 & 1.4680(6) & 1.4674(5) & 0.0006(11) & 1.4758(16) & 1.4602(15) & 0.0156(31) \\
S1$-$O4 & 1.5385(6) & 1.5372(5) & 0.0013(11) & 1.5577(16) & 1.5354(14) & 0.0223(30) \\

\end{tabular}
\end{ruledtabular}
\end{center}
\end{table}

\begin{table}[ht]
\caption{\label{tab:6} Diagonal thermal factors in \AA$^2$ units: $U_{\mathrm{H/D}x}(T)$ parallel to O$\cdots$O and $U_{\mathrm{H/D}z}(T)$ perpendicular to O$\cdots$O and parallel to $(a,b)$. }
\begin{center}
\begin{ruledtabular}
\begin{tabular}{lll}

$U_{\mathrm{H}x}(5) = 0.0351$ & $U_{\mathrm{H}x}(12) = 0.0326$ & $U_{\mathrm{H}x}(293) = 0.0932$\\
                 & $U_{\mathrm{D}x}(14) = 0.0937$ & $U_{\mathrm{D}x}(293) = 0.1117$\\
\hline
$U_{\mathrm{H}z}(5) = 0.0162$ & $U_{\mathrm{H}z}(12) = 0.0160$ & $U_{\mathrm{H}z}(293) = 0.0290$\\
                 & $U_{\mathrm{D}z}(14) = 0.0141$ & $U_{\mathrm{D}z}(293) = 0.0254$\\
\end{tabular}
\end{ruledtabular}
\end{center}
\end{table}

\begin{table}[ht]
\caption{\label{tab:7} Computed half-widths at half-height in \AA\ units of the probability densities: $\Delta_{\mathrm{H/D}x}(T)$ parallel to O$\cdots$O; $\Delta_{\mathrm{H/D}z}(T)$ perpendicular to O$\cdots$O and parallel to $(a,b)$; $\Delta_{\mathrm{H/D}y}(T)$ parallel to $c^*$. }
\begin{center}
\begin{ruledtabular}
\begin{tabular}{lll}

$\Delta_\mathrm{Hx}(5) = 0.156$ & $\Delta_\mathrm{Hz}(5) = 0.106$ & $\Delta_\mathrm{Hy}(5) = 0.095$ \\
$\Delta_\mathrm{Hx}(12) = 0.150$ & $\Delta_\mathrm{Hz}(12) = 0.105$ & $\Delta_\mathrm{Hy}(12) = 0.111$ \\
$\Delta_\mathrm{Hx}(293) = 0.254$ & $\Delta_\mathrm{Hz}(293) = 0.145$ & $\Delta_\mathrm{Hy}(293) = 0.142$ \\
\hline
$\Delta_\mathrm{Dx}(14) = 0.255$ & $\Delta_\mathrm{Dz}(14) = 0.099$ & $\Delta_\mathrm{Dy}(14) = 0.088$ \\
$\Delta_\mathrm{Dx}(293) = 0.278$ & $\Delta_\mathrm{Dz}(293) = 0.133$ & $\Delta_\mathrm{Dy}(293) = 0.124$ \\

\end{tabular}
\end{ruledtabular}
\end{center}
\end{table}

\begin{table}[p]
\caption{\label{tab:8} Tentative assignment scheme of the OH stretching transitions observed with inelastic neutron scattering at 20 K, from Ref. [\onlinecite{FLTK}], and comparison with the calculated transitions for a bare proton in an infinite square well whose width is $a = 0.47$ \AA}
\begin{center}
\begin{ruledtabular}
\begin{tabular}{lllllll}

 & & $1-2$ & $1-3$ & $1-4$ & $1-5$ & $1-6$ \\
\hline
K$_3$H(SO$_4)_2$ & Obs. (\cm) & $57\pm3$ & $145\pm5$ & $280\pm10$ & $445\pm15$ & $680\pm20$ \\
                 & Calc. (\cm) & $57\pm3$ & $152\pm8$ & $285\pm15$  & $456\pm24$ & $665\pm 35$ \\
\hline
Rb$_3$H(SO$_4)_2$ & Obs. (\cm) & $44\pm$? & $145\pm5$ & $260\pm10$ & $440\pm15$ & $680\pm20$ \\
%                 & $E_1$ (\cm) & $14.7\pm$? & $18.1\pm0.6$ & $17.3\pm0.7$  & $18.3\pm0.6$ & $19.5\pm0.6$ \\

\end{tabular}
\end{ruledtabular}
\end{center}
\end{table}

\clearpage

\section*{list of tables}

\contentsline {table}{\numberline {I}{\ignorespaces Neutron single crystal diffraction data and structure refinement for K$_3$H(SO$_4)_2$ and K$_3$D(SO$_4)_2$. $\lambda $ = 0.8305 \r A. Space groups monoclinic $A 2/a$ with Z = 8. The criterion for used reflections was I $>$ 3$\sigma $(I). The variance for the last digit is given in parentheses. Refinement on F.}}{19}{}
\contentsline {table}{\numberline {II}{\ignorespaces Atomic positions and isotropic temperature factors for K$_3$H(SO$_4)_2$ at 5 K (first lines), 12 K (second lines) and 293 K (third lines).}}{20}{}
\contentsline {table}{\numberline {III}{\ignorespaces Atomic positions and isotropic temperature factors for K$_3$D(SO$_4)_2$ at 14 K (first lines) 293 K (second lines).}}{21}{}
\contentsline {table}{\numberline {IV}{\ignorespaces Thermal parameters in \r A$^2$ units for K$_3$H(SO$_4)_2$ at 5 K (first lines), 12 K (second lines) and 293 K (third lines). The thermal parameters account for the variation of the contribution of each atom to Bragg's peak intensities through the thermal factor $T^{at}$ depending on the reciprocal lattice parameters $a^*$, $b^*$, $c^*$, and unit cell indexes in reciprocal space $h, k, l$, as $T^{at} = \mathop {\mathgroup \symoperators exp}\nolimits [ -2\pi ^2 (U_{11}^{at}h^2a^{*2} + U_{22}^{at}k^2b^{*2} + U_{33}^{at}l^2c^{*2} + 2U_{12}^{at}ha^*kb^* + 2U_{23}^{at}kb^*lc^* + U_{31}^{at}lc^*ha^* ) ]$.}}{22}{}
\contentsline {table}{\numberline {V}{\ignorespaces Thermal parameters in \r A$^2$ units for K$_3$D(SO$_4)_2$ at 14 K (first lines) and 293 K (second lines). The thermal parameters account for the variation of the contribution of each atom to Bragg's peak intensities through the thermal factor $T^{at}$ depending on the reciprocal lattice parameters $a^*$, $b^*$, $c^*$, and unit cell indexes in reciprocal space $h, k, l$, as $T^{at} = \mathop {\mathgroup \symoperators exp}\nolimits [ -2\pi ^2 (U_{11}^{at}h^2a^{*2} + U_{22}^{at}k^2b^{*2} + U_{33}^{at}l^2c^{*2} + 2U_{12}^{at}ha^*kb^* + 2U_{23}^{at}kb^*lc^* + U_{31}^{at}lc^*ha^* ) ]$.}}{23}{}
\contentsline {table}{\numberline {VI}{\ignorespaces Bond lengths in \r A\ units of the SO$_4$ entities and their expansion $\Delta $ with temperature. }}{24}{}
\contentsline {table}{\numberline {VII}{\ignorespaces Diagonal thermal factors in \r A$^2$ units: $U_{\mathrm {H/D}x}(T)$ parallel to O$\cdots $O and $U_{\mathrm {H/D}z}(T)$ perpendicular to O$\cdots $O and parallel to $(a,b)$. }}{24}{}
\contentsline {table}{\numberline {VIII}{\ignorespaces Computed half-widths at half-height in \r A\ units of the probability densities: $\Delta _{\mathrm {H/D}x}(T)$ parallel to O$\cdots $O; $\Delta _{\mathrm {H/D}z}(T)$ perpendicular to O$\cdots $O and parallel to $(a,b)$; $\Delta _{\mathrm {H/D}y}(T)$ parallel to $c^*$. }}{24}{}
\contentsline {table}{\numberline {IX}{\ignorespaces Tentative assignment scheme of the OH stretching transitions observed with inelastic neutron scattering at 20 K, from Ref. [\onlinecite {FLTK}], and comparison with the calculated transitions for a bare proton in an infinite square well whose width is $a = 0.47$ \r A}}{25}{}

\section*{list of figures}

\contentsline {figure}{\numberline {1}{\ignorespaces Projections of dimer entities onto $(a,b)$ planes. Thermal ellipsoids represent $50 \%$ of the probability density for nuclei. Sulfur is yellow, oxygen is red, potassium is black, hydrogen or deuterium are grey. }}{5}{}
\contentsline {figure}{\numberline {2}{\ignorespaces INS spectra of K$_3$H(SO$_4)_2$ and Rb$_3$H(SO$_4)_2$ crystal powders at 20 K, after ref. [\onlinecite{FLTK}]. The bands at $\approx 600$ cm$^{-1}$\ were assigned to SO$_4$ entities. }}{6}{}
\contentsline {figure}{\numberline {3}{\ignorespaces Eigenstates of the square-well along the OH stretching coordinate}}{7}{}
\contentsline {figure}{\numberline {4}{\ignorespaces Solid lines: Computed probability densities along the OH stretching coordinate at 5 K (bottom) and 293 K (top) for a rigid H$_{1/2}-$H$_{1/2}$ entity. The length is $l_\mathrm {H} = 0.16$ \r A\ and the center of mass experiences a square-well whose width is $0.47$ \r A. Dashed line: probability density for each half proton. $+$: Gaussian profile whose variance is $U_{x}^\mathrm {H}(T)/2$ (see Table VII{}{}{}\hbox {}). }}{8}{}
\contentsline {figure}{\numberline {5}{\ignorespaces Solid lines: Computed probability densities along the OD stretching coordinate at 14 K, 85 K, and 293 K for a rigid D$_{L1/2}-$D$_{R1/2}$ entity. The length is $ l_\mathrm {D} = 0.25$ \r A\ and the center of mass experiences a box whose width is $0.47$ \r A. Dashed line: contributions of the $n = 1$ and $n = 2$ states to the probability density. $+$: Gaussian profile whose variance is $U_{x}^\mathrm {D}(T)/2$ (Table VII{}{}{}\hbox {}). }}{9}{}
\contentsline {figure}{\numberline {6}{\ignorespaces ($a$) The indiscernibility degree of the ground state, $\mathcal {I}_1(l)$, as a function of the separation of the split-wave. $l_\mathrm {H} = 0.16$ \r A, $\mathcal {I}_1 = 0.60$ for K$_3$H(SO$_4)_2$. $l_\mathrm {D} = 0.25$ \r A, $\mathcal {I}_1 = 0.27$ for K$_3$D(SO$_4)_2$. $l_{di} = 0.183$ \r A, $\mathcal {I}_{1di} = 0.50$ at the discernible-indiscernible boundary. ($b$) The transition parameter $\mathcal {IP} = \mathcal {I}_1$ if $l \leq l_{di}$, $\mathcal {IP} = 1$ if $l \geq l_{di}$ . }}{12}{}
\contentsline {figure}{\numberline {7}{\ignorespaces Temperature laws for the indiscernibility parameter driving the phase transition of K$_3$D$_{(1-\rho )}$H$_\rho $(SO$_4)_2$, according to eqs (8{}{}{}\hbox {}) and (10{}{}{}\hbox {}). Dash: the dividing line between discernibility (below) and indiscernibility (above) domains. }}{13}{}
\contentsline {figure}{\numberline {8}{\ignorespaces Solid line: Temperature of the discernible-indiscernible transition for K$_3$D$_{(1-\rho )}$H$_\rho $(SO$_4)_2$, according to eq. (11{}{}{}\hbox {}). $\bullet $: experimental data (dielectric measurements) estimated from digitized Fig. 3 in Ref. [\onlinecite{MTNSK}]. DIB: discernible-indiscernible boundary. $l_\rho = (1-\rho )l_\mathrm {D} + \rho l_\mathrm {H}$ is the effective separation length for the components of the wavefunction.}}{14}{}
\contentsline {figure}{\numberline {9}{\ignorespaces Solid: specific heat jump for K$_3$D$_{(1-\rho )}$H$_\rho $(SO$_4)_2$, according to eq. (13{}{}{}\hbox {}). Dot dash: the Ising model, as shown in Ref. [\onlinecite{MTNSK}]. $\bullet $: experimental data estimated from digitized Fig. 4 in Ref. [\onlinecite{MTNSK}]. DIB: discernible-indiscernible boundary. $l_\rho = (1-\rho )l_\mathrm {D} + \rho l_\mathrm {H}$ is the effective separation length for the components of the wavefunction.}}{15}{}

\end{document}